\documentclass[conference]{IEEEtran}
\IEEEoverridecommandlockouts

\usepackage{textcomp}
\usepackage{xcolor}
\usepackage{booktabs}       
\usepackage{makecell}       
\usepackage{array}          
\usepackage{url}            
\usepackage[style=numeric]{biblatex}

\def\BibTeX{{\rm B\kern-.05em{\sc i\kern-.025em b}\kern-.08em
    T\kern-.1667em\lower.7ex\hbox{E}\kern-.125emX}}

\begin{document}
\title{Institutional Research Computing Capabilities in Australia: 2024}

\author{
\IEEEauthorblockN{1\textsuperscript{st} Slava Kitaeff}
\IEEEauthorblockA{\textit{The University of New South Wales}\\
\textit{Monash University}\\
Melbourne, Australia\\
s.kitaeff@unsw.edu.au, slava.kitaeff@monash.edu}
\and
\IEEEauthorblockN{2\textsuperscript{nd} Luc Betbeder-Matibet}
\IEEEauthorblockA{\textit{The University of New South Wales}\\
Sydney, Australia\\
luc@unsw.edu.au}
\and
\IEEEauthorblockN{3\textsuperscript{rd} Jake Carroll}
\IEEEauthorblockA{\textit{The University of Queensland}\\
Brisbane, Australia\\
jake.carroll@uq.edu.au}
\and
\IEEEauthorblockN{4\textsuperscript{th} Stephen Giugni}
\IEEEauthorblockA{\textit{The University of Melbourne}\\
Melbourne, Australia\\
stephen.giugni@unimelb.edu.au}
\\
\and
\IEEEauthorblockN{5\textsuperscript{th} David Abramson}
\IEEEauthorblockA{\textit{The University of Queensland}\\
\textit{Monash University}\\
Brisbane, Australia\\
david.abramson@uq.edu.au}
\and
\IEEEauthorblockN{6\textsuperscript{th} John Zaitseff}
\IEEEauthorblockA{\textit{The University of New South Wales}\\
Sydney, Australia\\
j.zaitseff@unsw.edu.au}
\and
\IEEEauthorblockN{7\textsuperscript{th} Sarah Walters}
\IEEEauthorblockA{\textit{The University of Queensland}\\
Brisbane, Australia\\
sarah.walters@uq.edu.au}
\and
\IEEEauthorblockN{7\textsuperscript{th} David Powell}
\IEEEauthorblockA{\textit{Monash University}\\
Melbourne, Australia\\
david.powell@monash.edu}
\and
\IEEEauthorblockN{8\textsuperscript{th} Chris Bording}
\IEEEauthorblockA{\textit{The University of Western Australia}\\
Perth, Australia\\
chris.bording@uwa.edu.au}
\\
\and
\IEEEauthorblockN{9\textsuperscript{th} Trung Nguyen}
\IEEEauthorblockA{\textit{The Commonwealth Scientific}\\\textit{\& Industrial Research Organisation}\\
Melbourne, Australia\\
trung.nguyen@csiro.au}
\and
\IEEEauthorblockN{10\textsuperscript{th} Angus Macoustra}
\IEEEauthorblockA{\textit{The Commonwealth Scientific}\\\textit{\& Industrial Research Organisation}\\
Melbourne, Australia\\
Angus.Macoustra@csiro.au}
\and
\IEEEauthorblockN{11\textsuperscript{th} Fabien Voisin}
\IEEEauthorblockA{\textit{The University of Adelaide}\\
Adelaide, Australia\\
fabien.voisin@adelaide.edu.au}
\\
\and
\IEEEauthorblockN{12\textsuperscript{th} Bowen Chen}
\IEEEauthorblockA{\textit{The University of Adelaide}\\
Adelaide, Australia\\
bowen.chen@adelaide.edu.au}
\and
\IEEEauthorblockN{13\textsuperscript{th} Jarrod Hurley}
\IEEEauthorblockA{\textit{Swinburne University}\\
Melbourne, Australia\\
jhurley@swin.edu.au}
}
 
\maketitle

\begin{abstract}
Institutional research computing infrastructure plays a vital role in Australia's research ecosystem, complementing and extending national-level facilities. This paper presents an analysis of research computing capabilities across Australian universities and research organisations, examining how institutional infrastructure supports research excellence through localised compute resources, specialised hardware, and cluster solutions. Our study reveals that institutional computing resources of nearly 112,258 CPU cores and 2,241 GPUs serve as essential bridges between desktop computing and national facilities for over 6,000 researchers, enabling research workflows that span from development to large-scale computations. We estimate the total replacement value of this infrastructure to be approximately \$144M AUD. Based on detailed infrastructure data provided by research computing facilities across multiple institutions, we identify key patterns in infrastructure deployment, utilisation metrics, and strategic alignment with research priorities. Our findings demonstrate that institutional computing resources not only provide critical support for data-intensive research but also facilitate training and higher-degree research student projects, enable prototyping and development, and ensure data sovereignty compliance when necessary. The analysis shows how these facilities leverage national infrastructure investments while addressing institution-specific needs that cannot be met by national facilities alone. We present evidence that strategic investment in institutional research computing capabilities yields significant returns through increased research productivity, enhanced graduate training, and improved research outcomes. This study provides valuable insights for research organisations planning their computing infrastructure strategies and highlights the importance of maintaining robust institutional computing capabilities alongside national facilities.
\end{abstract}


\section{Introduction}
Research computing has become indispensable across modern scientific disciplines, from genomics to artificial intelligence. While national facilities like the National Computational Infrastructure (NCI) \cite{NCI} and Pawsey Supercomputing Centre \cite{Pawsey} provide peak computing capabilities, institutional research computing infrastructure forms a critical layer in Australia's research ecosystem.

This institutional layer has evolved organically to address the specific needs of diverse research communities. It serves multiple essential functions: providing accessible resources for methodology development and testing, supporting specialised workflows that may not align with national facility requirements, enabling research software development, and building computational expertise among researchers. The geographical distribution of research activities across Australia and the growing computational demands of contemporary research methods further emphasise the importance of robust institutional computing capabilities.
Despite their significance, there has been no comprehensive analysis of institutional research computing infrastructure across Australia until now. This study examines computing capabilities at nine major institutions: UNSW Sydney, Monash University, University of Queensland, University of Melbourne, CSIRO, University of Adelaide, University of Western Australia, University of Sydney, and Swinburne University. Together, these institutions represent a substantial portion of Australia's research computing landscape, supporting thousands of researchers across diverse disciplines.

The timing of this analysis is particularly relevant given several concurrent trends reshaping the computational research landscape. Artificial intelligence has emerged as a mainstream research tool, driving unprecedented demand for GPU computing resources. Data-intensive methodologies are becoming prevalent across disciplines, while interactive computing and web-based interfaces are transforming how researchers interact with computational resources. These developments present both opportunities and challenges for institutional infrastructure planning and operational models.

Our study examines not only technical specifications but also operational approaches, user communities, and strategic directions. By analysing commonalities and variations across institutions, we provide insights into effective strategies for delivering research computing capabilities. This comprehensive view considers infrastructure scale, technology choices, support models, and funding approaches to understand both current capabilities and future requirements.
Understanding this landscape is valuable for multiple stakeholders. Research organisations gain benchmarks and insights for infrastructure planning. Funding bodies receive evidence of institutional investments' scale and impact. Researchers benefit from greater awareness of available resources and collaboration opportunities. Policymakers gain understanding of these facilities' critical role in maintaining research competitiveness.

The findings will inform strategic discussions about the future development of research computing capabilities in Australia. As computational methods become increasingly central to research excellence, maintaining robust institutional computing infrastructure grows more critical. This analysis provides a foundation for strategic planning and policy development to ensure Australian researchers can access the computational resources they need for world-class research.

\section{Methodology}

\subsection{Data Collection}

This study employed a structured data collection approach focusing on eight major Australian institutional level research computing facilities, also often referred to as Tier-2, in contrast with the two national level HPC facilities - National Computing Infrastructure (NCI) and Pawsey Supercomputing Centre, also refereed as Tier-1 facilities. Data was collected through a standardised template as part of the Tier 2 HPC community Birds of a Feather (BoF) session at eResearch Australasia 2024. The following facilities were included in this study:
\begin{itemize}
    \item{Katana cluster (UNSW Sydney) \cite{UNSW}}
    \item{MASSIVE M3 cluster (Monash University) \cite{Monash}}
    \item{Bunya cluster (University of Queensland) \cite{UQ}}
    \item{Spartan cluster (University of Melbourne) \cite{UMel}}
    \item{Petrichor cluster (CSIRO) \cite{CSIRO-Petrichor}}
    \item{Virga cluster (CSIRO) \cite{CSIRO-Virga}}
    \item{Phoenix  cluster (University of Adelaide) \cite{UoA}}
    \item{Kaya  cluster (University of Western Australia) \cite{UWA}}
    \item{Artemis (University of Sydney) \cite{USyd}}
    \item{OzSTAR and Ngarrgu Tindebeek clusters (Swinburne University) \cite{USwin}}
\end{itemize}

\subsubsection {Data Categories}
The data collection template focused on four primary areas:

\begin{itemize}
    \item{CPU and GPU resources, including core or GPU count, number of nodes, architectures and configurations, network infrastructure and interconnects, age, and backup strategies}
    \item{Operational Data, including staff numbers, funding models, data centre arrangements, system software and scheduler choices}
    \item{Usage statistics, including active user counts, disciplinary distribution, application profiles, resource utilisation patterns}
    \item{Strategic information, such as current challenges, future plans, recent changes, and lessons learned}
\end{itemize}

\subsection{Analysis Framework}

The data collection and analysis focused on four primary areas: infrastructure metrics (compute, storage, and network resources), operational data (staffing, funding, and facilities), usage statistics (user demographics and applications), and strategic information (challenges and future plans). To enable meaningful comparisons, we normalised metrics across institutions, standardising compute capacity to core counts, storage to petabyte scale, and user metrics to active users within a 12-month period.

\subsection{Limitations}

Several limitations should be noted. The study focuses on computational cluster technologies, excluding cloud computing capabilities that form an increasingly important part of the research computing landscape. Data currency represents a snapshot as of October 2024, with some variation in reporting approaches across institutions, particularly in storage capacity reporting (physical vs usable). Additionally, local institutional priorities and constraints influencing infrastructure choices are not fully captured in the quantitative metrics.

\section{Current Landscape Analysis}

The institutional research computing landscape in Australia represents a substantial national capability, complementing national facilities. Our analysis reveals a combined computing power exceeding 96,400 CPU cores distributed across eight major institutions, supporting diverse research workflows from exploratory development to production-scale computation.

\subsection{Computing Resources}

\subsubsection{CPU Infrastructure}
The analysis of CPU infrastructure across Australian institutions reveals distinct approaches to resource deployment based on research needs, institutional size, and funding models. Table \ref{tab:cpu-infrastructure} provides an overview of the key metrics for each institution.

\begin{table*}[htbp]
\centering
\caption{Institutional CPU Infrastructure Overview}
\label{tab:cpu-infrastructure}
\begin{tabular}{@{}lrrllr@{}}
\toprule
Institution & Total Cores & Nodes & Architecture Distribution & Avg Age (years)\\
\midrule
UNSW (Katana) & 7,384 & 165 & 97 Cascade Lake, 28 Skylake, 25 Ice Lake, AMD Zen 3/4 & 2.5\\
Monash (MASSIVE M3) & 14,312 & -- & Intel, AMD & 1-6\\
UQ (Bunya) & 14,800 & 145 & 57\% AMD Milan, 40\% AMD Genoa, 3\% Sapphire Rapids & 0.1-3\\
UMelb (Spartan) & 14,200 & -- & 50\% Cascade Lake, 50\% Sapphire Rapids & 2.5\\
CSIRO (Petrichor) & 25,000+ & 400+ & AMD Milan & 3\\
UoA (Phoenix) & 11,520 & 160 & Icelake & 1-2\\
UWA (Kaya) & 1,596 & 32 & AMD Genoa & $<$5\\
USyd (Artemis) & 7,588 & 195 & 3 different generations & 8\\
Swinburne (OzSTAR,\\Ngarrgu Tindebeek) & 15,860 & 300 & Intel Xeon Skylake \& AMD EPYC & 1-5\\
\bottomrule
\end{tabular}
\end{table*}

A clear tiered structure has emerged across the institutional landscape:
\begin{enumerate}
    \item Large-scale facilities (CSIRO, Monash, UQ, UMelb, Swinburne) each maintain over 14,000 cores to support broad research communities with diverse computational requirements.
    \item Medium-scale facilities (UNSW, UoA, USyd) operate between 7,000-12,000 cores, balancing resource availability with institutional priorities and funding constraints.
    \item Specialised facilities like UWA's Kaya provide focused support for specific research communities, with a smaller but targeted deployment of resources.
\end{enumerate}

Most facilities operate hardware less than 3 years old, reflecting relatively consistent refresh cycles across the sector. A notable shift toward AMD EPYC processors is evident in recent deployments, particularly at CSIRO, UQ, and UWA, while maintaining Intel presence for compatibility with established research codes. This architecture diversification suggests both increased market competition and growing sophistication in matching hardware to specific workload requirements.

Heterogeneous configurations predominate across all institutions, enabling support for diverse computational needs from high-throughput computing to memory-intensive applications. These varied approaches to CPU infrastructure demonstrate how institutional facilities have evolved to address local research priorities while maintaining flexibility to support emerging computational methods.

\subsubsection{GPU Infrastructure}

\begin{table*}[htbp]
\centering
\caption{Institutional GPU Infrastructure Overview}
\label{tab:gpu-infrastructure}
\begin{tabular}{@{}lrlll@{}}
\toprule
Institution & Total GPUs & GPU Types & GPU/Node & Primary Usage \\
\midrule
UNSW & 117 & A100, V100, L40S, H200 & 4-8 & AI/ML, Research \\
Monash & 490 & H100, A100, A40, A16, P4, T4, V100 & Various & Mixed workloads \\
UQ & 114 & A100, A16, L40/L40s, H100, MI210, MI300X & 2-8 & AI/ML, Research \\
UMelb & 408 & 80\% A100, 15\% H100, 5\% V100 & 4 & AI/ML, Research \\
CSIRO (Virga) & 444 & H100 & 4 & AI/ML, Research \\
UoA & 200 & A100 40GB SXM & 4 & AI/ML, Research \\
UWA & 38 & V100, A100, P100, AMD Mi210 & 2-4 & Mixed workloads \\
USyd & 108 & V100 & 4 & Research \\
Swinburne & 300 & P100, A100, V100 & 2-8 & AI/ML, Astronomy \\
\bottomrule
\end{tabular}
\end{table*}

As shown in Table \ref{tab:gpu-infrastructure}, the GPU infrastructure across Australian research institutions represents a substantial national investment in accelerated computing capabilities. The resource distribution reveals notable concentration patterns, with four institutions (Monash, CSIRO, UMelb, and Swinburne) hosting approximately 74\% of the national institutional GPU capacity. Monash leads with 22\% of the national share, followed by CSIRO (20\%), UMelb (18\%), and Swinburne (14\%). This concentration reflects both strategic investment decisions and institutional specialisation in computationally intensive research areas.

The GPU fleet exhibits strategic evolution across multiple dimensions. In terms of architecture and generation, the current deployment represents a multi-generational approach to GPU computing. Latest-generation NVIDIA H100 deployments at CSIRO, Monash, and UQ provide cutting-edge capabilities for the most demanding workloads. However, the deployment backbone consists of A100 GPUs, which offer an optimal balance of performance and availability. Legacy V100 GPUs continue to provide valuable compute capacity for workloads that don't require the latest architectural features. This generational diversity enables efficient resource allocation based on computational requirements while managing investment cycles.

Although NVIDIA hardware dominates the landscape, emerging diversification is evident in AMD deployments at UQ (MI300X) and UWA (Mi210). This nascent heterogeneity could have significant implications for future procurement strategies and software ecosystem development as the market evolves.

The convergence toward 4 GPUs per node has emerged as the preferred configuration across many institutions, representing an empirically-derived balance between computational density, power requirements, cooling capabilities, and interconnect bandwidth. However, several institutions (UNSW, UQ, Swinburne) maintain flexibility through varied configurations ranging from 2 to 8 GPUs per node, enabling support for specialised research workflows with different resource profiles.

The deployment patterns reflect both current research priorities and emerging computational trends in workload distribution. AI/ML applications dominate GPU usage across most institutions, driving significant investment in hardware optimised for these workloads. However, the infrastructure also supports diverse traditional GPU computing applications, including computational chemistry, molecular dynamics, and astronomy. The increasing provision of visualisation-focused GPUs (such as RTX6000 and A40 models) acknowledges the growing importance of real-time processing and visual analytics in modern research workflows.

Resource allocation metrics provide further insight into institutional strategies, with an average of 355 GPUs per 1000 active users across institutions. This ratio varies significantly between organisations, reflecting differences in research focus, user demographics, and allocation policies. The observed mix of general access and dedicated research group allocations highlights the ongoing challenge of balancing equitable access with the specific needs of GPU-intensive research programs.

These patterns collectively demonstrate a robust and strategically deployed national GPU computing capability. Institutions have made targeted investments aligned with both current research needs and emerging computational methods, while maintaining sufficient flexibility to adapt to evolving requirements. The diversity in deployment approaches reflects local research priorities while ensuring essential capabilities for common workloads remain available across the research ecosystem.

\subsection{Computational Storage Infrastructure}

\begin{table*}[htbp]
\centering
\caption{Storage Infrastructure Overview}
\label{tab:storage-infrastructure}
\begin{tabular}{@{}lllllr@{}}
\toprule
Institution & Scratch Storage & User/Group Storage & Backup Strategy & File System & Storage/Core Ratio (TB)\\
\midrule
UNSW & 9.5 PB & 8 TB & Home only & NFS & 1.29 \\
Monash & 1+2* PB & 9 PB & None & Lustre + Ceph & 0.84 \\
UQ & $>$2.7 PB & Storage as service** & Research Data Fabric & IB Connected & 0.18 \\
UMelb & 1 PB & 4.2 PB & Full backup & GPFS & 0.37 \\
CSIRO & 2.2 PB & NetApp datasets & NetApp -- full backup; scratch -- no backup & Lustre & 0.09 \\
UoA & 3.0 PB + 112 TB & Storage as service & Selective & GPFS & 0.27 \\
UWA & 221 TB & 1.8 PB & None & GPFS & 1.26 \\
USyd & 0.5 PB & 0.5 PB & Selective & Lustre & 0.13 \\
Swinburne & 1.5 PB & 20 PB & None & BeeGFS/Lustre & 0.25 \\
\bottomrule
\multicolumn{6}{l}{\small *1 PB flash swap plus 2 PB disc storage}\\
\multicolumn{6}{l}{\small **100 PB storage available for staging data in/out via an NFS mount}
\end{tabular}
\end{table*}

\subsubsection{Storage Infrastructure Patterns}

As detailed in Table \ref{tab:storage-infrastructure}, the storage infrastructure across Australian research computing facilities demonstrates a sophisticated multi-tier approach optimised for diverse computational workflows. With a combined scratch storage capacity exceeding 14.8 PB and substantial project storage allocations, institutions have developed architectures that balance performance, capacity, and data lifecycle management.

Parallel file systems form the cornerstone of high-performance storage solutions across the sector, with Lustre deployed at CSIRO, Monash, USyd, and Swinburne, while GPFS (IBM Spectrum Scale) serves UMelb, UoA, and UWA. These choices reflect the need for high bandwidth and IOPS to support intensive computational workloads across large CPU and GPU clusters. Representative implementations include Monash's and UQ's scratch space that combine flash and disk tiers served through Lustre, and UoA's 3 PB GPFS deployment. These systems deliver data transfer rates commensurate with the demanding requirements of modern GPU-accelerated computing and multi-node CPU workloads.

The storage-to-compute ratios reveal strategic differences in institutional approaches to data-
intensive computing. These ratios range from 0.09 TB to 1.29 TB per CPU core, with a median of approximately 0.24 TB per core across the surveyed institutions. UNSW and UWA represent the upper end of this spectrum (1.29 and 1.26 TB per core, respectively), suggesting infrastructure optimised for particularly data-intensive research workflows. In contrast, CSIRO's lower ratio (0.09 TB per core) indicates a focus on compute-intensive rather than data-intensive workloads. Swinburne's ratio of 0.25 TB per core closely aligns with the median, reflecting a balanced approach to storage provisioning.

A clear pattern emerges in the separation between high-performance scratch storage and longer-term project storage. Most institutions maintain distinct tiers with different performance characteristics, reliability guarantees, and retention policies. Scratch storage typically offers higher performance but without backup, while project storage provides more robust data protection through various backup strategies. This architectural approach allows institutions to optimise cost-performance trade-offs across the storage lifecycle.

Emerging storage technologies are being selectively incorporated to address specialised workload requirements. The deployment of flash storage tiers is particularly notable, with examples including UQ's 40 TB, Adelaide's 4 TB fileset in a 100 TB flash-based software mount point, and Monash's 1 PB flash component in scratch storage. These high-IOPS storage tiers are strategically positioned to accelerate application loading, metadata operations, and small-file workloads, demonstrating increased sophistication in matching storage technologies to specific computational patterns.

Operational metrics indicate systems functioning at or near capacity limits. Several institutions report utilisation rates exceeding 80\% across their storage infrastructure. UNSW's scratch space at 79\% utilisation and USyd's at 100\% highlight the ongoing challenge of balancing storage provisioning against resource constraints. This utilisation pattern underscores the tension between providing generous allocations to support data-intensive research and managing the significant costs associated with large-scale storage infrastructure.

\subsubsection{Integration with National Facilities}

The relationship between institutional and national computing resources demonstrates sophisticated strategies for creating comprehensive computational ecosystems that serve researchers across multiple scales. These integration approaches vary based on geographical location, research priorities, and institutional partnerships.

Western Australian institutions have developed strong integration with the Pawsey Supercomputing Centre. CSIRO and the University of Western Australia participate in partner share arrangements that provide guaranteed access to Setonix and other Pawsey capabilities. This integration enables these institutions to focus their local infrastructure investments on complementary capabilities that address specific research needs. UWA's Kaya cluster provides essential development and testing capabilities that complement their Pawsey access, while CSIRO maintains substantial local infrastructure (Petrichor and Virga) alongside their Pawsey partnership. This dual-tier approach supports specialised research domains including radio astronomy, geology and mineral sciences, and energy research at Pawsey, while leveraging NCI Gadi \cite{Gadi} for earth systems sciences and climate modelling.

In contrast, eastern seaboard universities have developed different strategic approaches. UNSW, Monash University, and the University of Melbourne have made additional investments in NCI beyond standard merit-based access. This investment strategy provides guaranteed computing capacity at NCI while maintaining substantial local infrastructure for methodology development, specialised workflows, and time-sensitive research needs. The University of Melbourne's approach with Spartan exemplifies how local infrastructure can be optimised to complement NCI access, particularly for data-intensive workflows that benefit from processing proximity to institutional data sources.

These complementary approaches create a multi-tier computing environment that enhances research capability through specialised roles at each level:

\begin{enumerate}
    \item Local clusters serve as development platforms where researchers can prototype, test, and refine computational methods before scaling to national resources.
    \item Institutional facilities provide accessible computational capacity for production research that requires moderate scale but frequent access or specialised configurations.
    \item National facilities deliver peak computing capabilities for large-scale computations that exceed institutional resources.
\end{enumerate}

This tiered ecosystem influences storage and data management strategies as well. Many institutions have implemented data transfer nodes and high-speed network connections to facilitate efficient movement between local and national systems. This infrastructure allows researchers to maintain working datasets locally while accessing larger-scale storage capabilities at national facilities when needed.

The emergence of this complementary computational ecosystem represents a sophisticated response to the diverse needs of the research community. It provides flexible pathways from method development to production research while ensuring efficient utilisation of both local and national resources. This approach has proven particularly valuable for computationally intensive domains like climate modelling, genomics, and artificial intelligence, where research workflows may span multiple computational scales.

\subsection{User Communities}

\begin{table*}[htbp]
\centering
\caption{User Community and Application Analysis}
\label{tab:user-community}
\begin{tabular}{@{}lrrlll@{}}
\toprule
Institution & \makecell[lt]{Active\\Users} & Cores/Users & Top Disciplines & Core Applications & Usage Pattern \\
\midrule
UNSW & 682 & 10.9 & \makecell[lt]{Science (66.5\%), Res.Infra (16.2\%),\\Medicine (8.2\%)} & \makecell[lt]{R (20.0\%); Matlab (16.5\%);\\Python (9.9\%)} & \makecell[lt]{50 users $>10k$ jobs\\9 users $>100k$ jobs}\\
Monash & 897 & 15.9 & Data Science, Medicine, Engineering & Desktop, CryoSPARC, Jupyter & \makecell[lt]{Large data processing,\\I/ML workloads}\\
UQ & 1352 & 10.9 & AI/ML, Hypersonics, Genomics & \makecell[lt]{Python for AI,\\CFD solvers, Cryo-EM codes} & \makecell[lt]{Growing user diversity,\\Many new CLI users} \\
UMelb & 915 & 15.6 & Mech Eng, Computing, Bioinformatics & \makecell[lt]{Python, AI/ML, GROMACS, RStudio,\\Jupyter, CryoSPARC} & Mixed workload profile \\
CSIRO & 650+ & 38.5 & CFD, Climate & CFD codes, AI/ML tools, Climate models & \makecell[lt]{Research and operational\\loads}\\
UoA & 450* & 23.3 & AIML, Physics, ChemEng, Bioinformatics & \makecell[lt]{TensorFlow, Ansys, Matlab, Snakemake,\\Nextflow, CryoSPARC} & Mixed workload profile \\
UWA & 150 & 10.6 & Chemistry, Oceans, Engineering & Gaussian, Ansys, Abaqus & -- \\
USyd & 657 & 11.5 & Science, Engineering, Medicine & Python, R, Matlab & \makecell[lt]{6 users $>$1M CPUh;\\231 users $>$10k CPUh}\\
Swinburne & 328 & 37.4 & \makecell[lt]{Astronomy (70\%), Data Science (15\%),\\Physics (10\%)} & 
Python, TensorFlow, CASA & \makecell[lt]{Astronomy pipelines,\\ML/AI research}\\
\bottomrule
\multicolumn{6}{l}{\small *Average of reported range 400-500}
\end{tabular}
\end{table*}

User demographics presented in Table \ref{tab:user-community} reveal significant variation across institutions, from focused communities of around 150 researchers at UWA to large-scale deployments supporting over 1,350 users at UQ. Cores-per-user ratios range from 10.6 at UWA to 38.5 at CSIRO, reflecting different approaches to resource provisioning. CSIRO's higher ratio indicates emphasis on computationally intensive workloads, while universities typically provide more democratised access supporting broader research communities.

Usage patterns show strong disciplinary concentrations. At UNSW, Science faculty accounts for ~66\% of jobs on Katana cluster. Monash demonstrates strength in genomics and medicine, while UQ reports significant AI/ML and Hypersonics research. Swinburne shows a strong focus on astronomy (70\%), reflecting its support for the OzGrav Centre of Excellence. The emergence of AI/ML as a primary discipline at multiple institutions signals a shift in computational requirements.

The application landscape combines commonality in core tools with disciplinary specialisation. Python dominates data science and AI/ML workflows, while MATLAB maintains strong presence in engineering applications and R in statistical computing. Specialised applications like GROMACS, Gaussian, and CryoSPARC reflect sophisticated domain-specific requirements. The presence of CryoSPARC and Cryo-EM codes at multiple institutions indicates growing support for data-intensive scientific instrumentation.

Usage intensity shows notable stratification. At UNSW, while 682 users were active, only 9 users ran more than 100,000 jobs, with one user accounting for 19.2\% of all jobs. Similarly, USyd reports 6 users consuming over 1 million CPU hours each. This pattern highlights the challenge of supporting both power users and occasional users within the same infrastructure.

\begin{table*}[htbp]
\centering
\caption{Institutional Operational Characteristics}
\label{tab:operational-models}
\begin{tabular}{@{}lrrllll@{}}
\toprule
Institution & Computing Staff FTEs & Users/Staff & Funding Model & Data Center Type & Scheduler & Department Structure \\
\midrule
UNSW & 9 & 75.9 & Mixed & On-premise & OpenPBS & Compute, Data, Community \\
Monash & 9 & 99.7 & Central+Partners & Commercial & SLURM & Infrastructure, Development \\
UQ & 10 & 136.4 & Central+Partners & Commercial & SLURM & Core infra, RSE, Support\\
UMelb & 9 & 28.6 & Mixed & On-premise & SLURM & Platforms, Development \\
CSIRO & 8 & 81.2 & Central & Commercial & SLURM & Platforms, Services, Science\\
UoA & 4 & 112.5 & Central & On-premise & SLURM & HPC support, eResearch\\
UWA & 3 & 50.0 & Mixed & Commercial & SLURM & HPC team \\
USyd & 6 & 109.5 & Mixed & Managed & PBSPro & ICT/SIH/MSP integration \\
Swinburne & 5 & 65.6 & Central+OzGrav & On-premise & SLURM & eResearch, Sciences\\
\bottomrule
\end{tabular}
\end{table*}

Operational models summarised in Table \ref{tab:operational-models} demonstrate diverse approaches to service delivery. Staff numbers range from 3-4 FTE at UWA and UoA to 8-10 FTE at UNSW, Monash, UQ, and CSIRO, with Swinburne operating at 5 FTE. Team structures have evolved beyond traditional system administration to include research software engineers, data specialists, and community engagement roles. This evolution reflects growing recognition of comprehensive user support requirements.

Three primary funding patterns emerge: pure central funding (UoA, UQ), mixed models combining central and researcher contributions (UNSW, UMelb, UWA), and partnership approaches (Monash, Swinburne). Mixed funding models often link researcher contributions to priority access, balancing equitable access with intensive research program support. Swinburne's model, combining central funding with contributions from the OzGrav Centre of Excellence, represents a specialised variation of the partnership approach.

Data centre strategies show strong trend toward commercial solutions, with five institutions using commercial facilities or managed services. This shift reflects both increasing operational complexity and need for reliable, scalable infrastructure. On-premise operations at UNSW, UMelb, UoA, and Swinburne suggest continued value in maintaining institutional data centre capabilities for specific use cases.

The transition to more accessible computing interfaces is evident across institutions. UNSW, UQ, and UMelb report significant OpenOnDemand adoption, while Monash emphasises desktop services. This trend toward interactive computing suggests democratisation of research computing resources previously dominated by traditional HPC workflows.

\section{Discussion}

Our analysis of institutional research computing across Australia's leading universities and CSIRO reveals not just the current state of infrastructure but also strategic patterns in how research organisations are responding to evolving computational demands. Several key themes emerge from this comprehensive assessment that have implications for future planning and policy development.

\subsection{Strategic Value of Institutional Infrastructure}

The substantial national investment in institutional computing capability, valued at approximately \$144M AUD, represents a strategic recognition of the critical role these resources play in Australia's research ecosystem. Beyond their raw computational capacity, these facilities serve several essential functions that complement rather than duplicate national infrastructure.

The comparison between institutional and national computing resources provides valuable context for understanding Australia's research computing ecosystem. While Gadi and Setonix together deliver approximately 450,000 CPU cores and 3,000 GPUs to the research community, our analysis shows that institutional facilities collectively contribute roughly 112,000 CPU cores and 2,241 GPUs—equivalent to about 20\% of the national CPU capacity and 43\% of the national GPU resources. This substantial institutional contribution demonstrates that these facilities aren't merely supplementary to national capabilities but constitute a crucial complementary layer in the research infrastructure. The value proposition differs significantly: where national facilities excel at delivering peak computing power for large-scale projects through competitive merit-based allocations, institutional infrastructure provides accessible, responsive computing tailored to local research needs, methodology development, and specialised workflows. This complementary relationship creates a robust multi-tier computing ecosystem that strengthens Australia's overall research capabilities across the full spectrum of research activities, from exploratory analysis to large-scale production computing.

First, institutional facilities provide a crucial middle tier between desktop computing and national peak facilities, offering researchers the computing power required for methodology development and testing before pursuing allocations on highly contested national resources. This development pathway is particularly important for early-career researchers and graduate students who may lack the preliminary results needed to secure competitive national allocations.

Second, these facilities enable specialised workflows that may not align well with national facility environments due to software requirements, interactive computing needs, or data sovereignty considerations. The diversity of architectures, software environments, and access models across institutions serves as a form of distributed innovation in research computing approaches.

Third, these facilities play a critical role in building computational expertise within research communities. The relative accessibility of institutional resources compared to national facilities makes them ideal environments for training and skills development, creating pathways for researchers to progress from basic to advanced computational methods.

\subsection{Evolution Toward Research-Optimised Heterogeneity}

The infrastructure evolution across institutions demonstrates a shift away from homogeneous HPC deployments toward heterogeneous environments optimised for diverse research requirements. This trend manifests in several ways:

The growing architectural diversity in CPU choices, with AMD EPYC processors gaining significant presence alongside Intel offerings, enables institutions to target specific performance characteristics to research workloads. Similarly, while NVIDIA maintains dominance in the GPU space, the emergence of AMD options suggests future potential for greater ecosystem diversity.

Storage architectures have evolved beyond simple scratch/project dichotomies toward sophisticated multi-tier solutions with targeted performance characteristics. The emergence of flash tiers for high-IOPS workloads, parallel file systems for large-scale I/O, and integrated data management approaches demonstrates increasing sophistication in matching infrastructure to research workflows.

Perhaps most significantly, the shift toward interactive computing and web-based interfaces, exemplified by OpenOnDemand adoption at several institutions, represents an important evolution in making computational resources accessible to researchers with varying levels of technical expertise. This democratisation of access potentially expands the impact of institutional investments beyond traditional computational science domains.

\subsection{Resource Allocation Challenges}

Despite the substantial investments in institutional computing, our analysis reveals persistent tensions in resource allocation and sustainability. The concentration of GPU resources at a few institutions creates potential inequities in access to these increasingly important accelerated computing capabilities. Similarly, the consistently high storage utilisation rates (exceeding 80\% at several institutions) indicate ongoing challenges in meeting data-intensive research demands.

The varying cores-per-user and GPUs-per-user ratios across institutions suggest different philosophical approaches to resource democratisation versus targeted allocation. Institutions with higher ratios (like CSIRO) prioritise sufficient resources for intensive computational workflows, while those with lower ratios emphasise broader access. Each approach has merit, but these differences may impact research opportunities and outcomes across the sector.

Funding models demonstrate similar diversity, with some institutions relying primarily on central funding while others employ mixed models that link researcher contributions to priority access. While these mixed models may better align resource allocation with demonstrated needs, they also risk creating inequities based on research funding disparities across disciplines.

\subsection{Future Directions and Challenges}

Looking forward, several critical challenges and opportunities will shape the evolution of institutional research computing in Australia:

The rapid expansion of AI/ML research has created unprecedented demand for GPU computing that shows no signs of abating. Institutions will need to develop sustainable approaches to GPU provisioning, potentially including new funding models, shared facilities, or specialised national capabilities to complement institutional resources.

Integration between on-premises infrastructure, cloud computing, and national facilities will become increasingly important. The current complementary relationships between institutional and national resources provide a foundation, but more sophisticated integration of cloud capabilities could enhance flexibility and cost-effectiveness.

Power, cooling, and space constraints present growing challenges, particularly for GPU-intensive computing. These infrastructure limitations may drive increased adoption of commercial data center solutions, as already seen at several institutions.

Technical expertise and support staffing represent potential bottlenecks in maximising the value of computational investments. The relatively consistent staff-to-user ratios across institutions suggests empirically-derived sustainable support models, but maintaining adequate expertise across increasingly complex technology stacks remains challenging.

Finally, determining the appropriate metrics for assessing the impact and effectiveness of institutional computing investments remains difficult. Moving beyond simple utilisation statistics toward measures of research productivity, educational impact, and domain expansion would better capture the multifaceted value these resources provide.

\subsection{Toward a National Framework}

The findings of this study suggest potential benefits from greater coordination and knowledge sharing across institutional computing facilities to solve common challenges. Areas where national frameworks or collaborative approaches might prove beneficial include:

\begin{itemize}
    \item Development of standardised metrics and reporting frameworks for research computing capabilities
    \item Collaborative approaches to training and documentation that leverage specialised expertise across institutions
    \item Federated identity management approach to enable integration of institutional and national facilities
\end{itemize}

Such collaboration need not diminish the valuable diversity in approaches that has emerged organically to address institution-specific needs and research priorities. Rather, it could enhance the collective capability of Australia's research computing ecosystem while still preserving distinctive institutional strengths and specialisations.

\section{Summary}

This comprehensive analysis of institutional research computing capabilities across major Australian universities and CSIRO reveals a sophisticated and strategically deployed national infrastructure valued at approximately \$144M AUD. The combined capability encompasses over 112,258 CPU cores, 2,241 research GPUs, and more than 14.8 PB of high-performance storage capacity, supporting over 6,080 active researchers across diverse disciplines. Notably, these institutional resources constitute approximately 20\% of the national CPU capacity and 43\% of the national GPU resources when compared to the Gadi and Setonix national facilities, highlighting their substantial contribution to Australia's research computing ecosystem. This infrastructure plays a vital role, providing essential computational capacity that bridges the gap between desktop computing and national peak facilities while offering more accessible, responsive computing tailored to local research needs.

The analysis reveals clear evidence of strategic adaptation to changing research needs, with strong investment in AI-capable infrastructure. Institutions have developed diverse but effective approaches to service delivery, with funding models typically combining institutional and researcher contributions.

 This significant institutional contribution demonstrates that rather than being overshadowed by national capabilities, these facilities form an essential and complementary layer in the research ecosystem. Where national facilities excel at delivering peak computing power for large-scale projects through competitive merit-based allocation processes, institutional infrastructure provides accessible, responsive computing tailored to local research needs, methodology development, and specialised workflows. This complementary relationship creates a robust multi-tier computing ecosystem that strengthens Australia's overall research capabilities by ensuring computational resources are available across the full spectrum of research activities, from exploratory analysis to large-scale production computing.

As research becomes increasingly computational and data-intensive, institutional computing facilities will continue to grow in importance. This study provides valuable guidance for future development, highlighting both achievements and opportunities. Future work should focus on developing standardised metrics for assessing capabilities, reconsidering traditional HPC terminology to better reflect the distinct role of institutional computing, and understanding the integration of cloud computing within the research computing landscape. These insights will be crucial for maintaining Australia's research competitiveness and ensuring researchers have access to the computational resources they need.

\printbibliography
\end{document}